\documentclass[12pt]{article}

\usepackage{graphicx,epsf,amssymb,amsbsy,amsfonts,amssymb,amsmath}
\usepackage[dvips]{color}

\newif\ifdraft
\drafttrue
\newif\ifpreprint
\preprinttrue

\def\sect#1{section~{\ref{#1}}}
\def\fig#1{fig.~{\ref{#1}}}

\def\figs#1#2{figs.~{\ref{#1}} and {\ref{#2}}}

\def\spa#1.#2{\left\langle#1\,#2\right\rangle}
\def\spb#1.#2{\left[#1\,#2\right]}
\def\spash#1.#2{\spa{\smash{#1}}.{\smash{#2}}}
\def\spbsh#1.#2{\spb{\smash{#1}}.{\smash{#2}}}
\def\sand#1.#2.#3{%
\left\langle\smash{#1}{\vphantom1}^{-}\right|{#2}%
\left|\smash{#3}{\vphantom1}^{-}\right\rangle}
\def\sandpp#1.#2.#3{%
\left\langle\smash{#1}{\vphantom1}^{+}\right|{#2}%
\left|\smash{#3}{\vphantom1}^{+}\right\rangle}
\def\sandpm#1.#2.#3{%
\left\langle\smash{#1}{\vphantom1}^{+}\right|{#2}%
\left|\smash{#3}{\vphantom1}^{-}\right\rangle}
\def\sandmp#1.#2.#3{%
\left\langle\smash{#1}{\vphantom1}^{-}\right|{#2}%
\left|\smash{#3}{\vphantom1}^{+}\right\rangle}

\def\tree{{\rm tree}}
\def\pol{\varepsilon}

\def\eps{\epsilon}

\def\nn{\nonumber}

\def\eqn#1{eq.~(\ref{#1})}

\def\NeqFour{{{\cal N}=4}}
\def\Neqeight{{{\cal N}=8}}
\def\NeqEight{{{\cal N}=8}}

\def\be{\begin{equation}}
\def\ee{\end{equation}}
\def\bea{\begin{eqnarray}}
\def\eea{\end{eqnarray}}
\def\ba{\begin{eqnarray}}
\def\ea{\end{eqnarray}}

\def\tree{{\rm tree}}

\newbox\charbox
\newbox\slabox
\def\s#1{{      
        \setbox\charbox=\hbox{$#1$}
        \setbox\slabox=\hbox{$/$}
        \dimen\charbox=\ht\slabox
        \advance\dimen\charbox by -\dp\slabox
        \advance\dimen\charbox by -\ht\charbox
        \advance\dimen\charbox by \dp\charbox
        \divide\dimen\charbox by 2
        \raise-\dimen\charbox\hbox to \wd\charbox{\hss/\hss}
        \llap{$#1$} }}

\def\subtractfour#1{\ifthenelse{#1=5}{1}{\ifthenelse{#1=6}{2}
{\ifthenelse{#1=7}{3}{\ifthenelse{#1=8}{4}{\ifthenelse{#1=9}{5}
{\ifthenelse{#1=10}{6}{\ifthenelse{#1=11}{7}{\ifthenelse{#1=12}{8}
{\ifthenelse{#1=13}{9}{\ifthenelse{#1=14}{10}{}}}}}}}}}}}

\begin{document}

\ifpreprint
 \hfill UCLA/08/TEP/40 
\fi

\vskip .3 cm 
\centerline{\Large \bf Progress on Ultraviolet Finiteness of Supergravity}

\vskip .3 cm 
\begin{center}
Z.~Bern\footnote{Presenter at International School of Subnuclear Physics, 
46th Course, Erice Sicily, August 29-September 7, 2008.},
J.~ J.~M.~Carrasco\footnote{Scientific secretary.} and H.~Johansson$^2$
\end{center}

\begin{center}
Department of Physics and Astronomy\\
UCLA, Los Angeles, CA\\
90095-1547, USA 
\end{center}

\date{December, 2008}

\begin{abstract}
In this lecture we summarize recent calculations pointing to the
possible ultraviolet finiteness of $\NeqEight$ supergravity in four
dimensions.  We outline the modern unitarity method, which enables
multiloop calculations in this theory and allows us to exploit a
remarkable relation between tree-level gravity and gauge-theory
amplitudes.  We also describe a link between observed cancellations at
loop level and improved behavior of tree-level amplitudes under large
complex deformations of momenta.
\end{abstract}



\section{Introduction}
\label{Introduction}

For over 25 years the prevailing wisdom has been that it is impossible
to construct a perturbatively ultraviolet finite point-like quantum
field theory of gravity in four dimensions (see {\it e.g.}
refs.~\cite{Supergravity}).  In this lecture we describe recent
concrete calculations that call into question this belief. 

Of all unitary quantum gravity field theories, maximally
supersymmetric $\NeqEight$ supergravity~\cite{CremmerJuliaScherk} is
the most promising one to investigate for possible ultraviolet
finiteness. Its high degree of supersymmetry suggests that it has the
best ultraviolet properties of any gravity field theory with two
derivative couplings.  Moreover, with the modern unitarity
method~\cite{UnitarityMethod, DDimUnitarity, GeneralizedUnitarity,
BDDPR, BCFUnitarity, FiveLoop, LeadingSingularity}, the high degree of
supersymmetry can be exploited to greatly simplify calculations.  In
fact, the striking simplicity of the theory led to the recent
suggestion that $\NeqEight$ supergravity may in a sense be the
simplest quantum field theory~\cite{AHCKGravity}.  The unitarity
method allows us to exploit a remarkable relation between gravity and
gauge-theory tree amplitudes~\cite{KLT, Grant, FreitasGravity},
allowing us to map gravity calculations into algebraically simpler
gauge-theory calculations.

In a classic paper 't~Hooft and Veltman showed that gravity coupled to
matter generically diverges at one loop in four
dimensions~\cite{tHooftVeltmanGravity, DeserMatter}.  Due to the
dimensionful nature of the coupling, the divergences cannot be
absorbed by a redefinition of the original parameters of the
Lagrangian, rendering the theory non-renormalizable.  Pure Einstein
gravity does not possess a viable counterterm at one loop, delaying
the divergence to two loops~\cite{tHooftVeltmanGravity, Kallosh74,
vanNWu}.  The two-loop divergence of pure Einstein gravity was
established by Goroff and Sagnotti and by van~de~Ven, through direct
computation~\cite{GoroffSagnotti,vandeVen}.

Over the years supersymmetry has been studied extensively as a
mechanism for delaying the onset of divergences in gravity theories.
No supergravity theory can diverge until at least three
loops~\cite{GrisaruTomboulis,DeserKayStelle,Supergravity}. With
additional assumptions this can be delayed
further~\cite{HoweStelleNew,Berkovits, GreenII, KellyPrivate,
HoweStelleRecent}, perhaps even to nine loops.  However, all
studies to date based on supersymmetry considerations alone point to
protection against divergences from supersymmetry failing at some loop
order. An additional mechanism is needed to prevent divergences.

Here we review direct evidence at three
loops~\cite{GravityThree,CompactThree} that such a mechanism does
indeed exist in gravity theories. It is striking that {\it
all} explicit complete calculations of $\NeqEight$ supergravity
scattering amplitudes~\cite{BDDPR, OneloopMHVGravity, NoTriangle,
NoTriangleB, GravityThree, NoTri, BjerrumVanhove, NoTriangleProof,
AHCKGravity, CompactThree} find a power counting identical to that of
$\NeqFour$ super-Yang-Mills theory, which is known to be finite in
four dimensions~\cite{West, Mandelstam, HoweStelleYangMills}.
Interestingly, M theory has also been used to argue for the finiteness
of $\NeqEight$ supergravity~\cite{DualityArguments}, though issues
with decoupling towers of massive states~\cite{GOS} may alter these
conclusions.

Here we focus only on the possible order-by-order finiteness of the
perturbative series of $\NeqEight$ supergravity, leaving aside
non-perturbative issues or how one might go about constructing a
realistic ultraviolet finite theory.  We are interested in
perturbative ultraviolet finiteness of gravity because the existence
of cancellations sufficient to render the theory finite would imply a
new symmetry or dynamical mechanism to enforce these.  We can expect
that a proper understanding of the mechanism behind such cancellations
will have a profound impact on our understanding of gravity.

\section{Quantum gravity}

\begin{figure}[t]
\centerline{\epsfxsize 2. truein \epsfbox{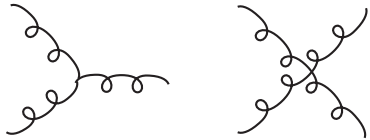}}
\caption[a]{\small Gauge theories have three- and four-point vertices
in a Feynman diagrammatic description.}
\label{YMVertFigure}
\end{figure}

\begin{figure}[t]
\centerline{\epsfxsize 5.0 truein \epsfbox{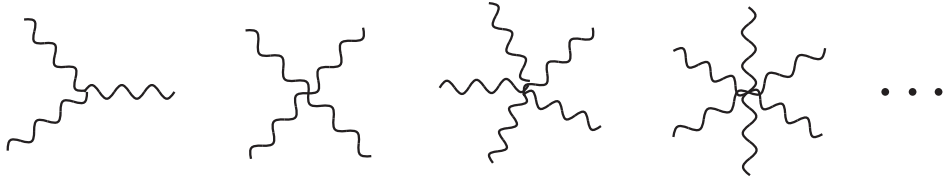}}
\caption[a]{\small Gravity theories have an infinite number of 
higher-point contact interactions in a Feynman diagrammatic description.}
\label{GravVertFigure}
\end{figure}

\subsection{Feynman diagram approach}

Before turning to the on-shell methods used in modern calculations, it
is helpful to first survey gravity Feynman diagrams.  We compare the
Feynman diagrams of gravity to those of gauge theory.  Consider the
Einstein-Hilbert and Yang-Mills Lagrangians,
\begin{equation}
{\cal L}_{\rm YM} = - \frac{1}{4} F^a_{\mu\nu} F^{a\, \mu\nu}
 \,, \hskip 2 cm  
{\cal L}_{\rm EH} = \frac{2}{\kappa^2}\sqrt{-g} R \,.
\end{equation}
Following standard Feynman diagrammatic methods we gauge fix and then
expand the Lagrangians in a set of vertices.  As illustrated in
\figs{YMVertFigure}{GravVertFigure}, with standard gauge choices for
gauge theory there are three- and four-point interactions, while for
gravity there are an infinite number of contact interactions.  (As we
shall discuss below, all interactions beyond three points are, in
fact, unnecessary when using on-shell methods.)  Perhaps more
striking than the infinite number of interactions is the complexity
of these interactions.  Consider, for example, the three-graviton
interaction.  In standard de Donder gauge, the three-graviton vertex
is of the form,
\begin{eqnarray}
&& \hskip -.7 cm 
G_{3\mu\alpha,\nu\beta,\sigma\gamma}(k_1,k_2,k_3)\nn \\ 
&&\null \hskip .5 cm 
 =  i \frac{\kappa}{2}
 {\rm Sym}\Bigl[ - \frac{1}{2} P_3(k_1\cdot k_2\eta_{\mu\alpha}\eta_{\nu\beta}
\eta_{\sigma\gamma}) - \frac{1}{2} P_6 (k_{1\nu}k_{1\beta}
\eta_{\mu\alpha}\eta_{\sigma\gamma}) + \frac{1}{2} 
P_3 (k_1\cdot k_2 \eta_{\mu\nu}\eta_{\alpha\beta}\eta_{\sigma\gamma}) \nn\\
&&\null \hskip 1.8cm 
+ P_6(k_1\cdot k_2 \eta_{\mu\alpha}\eta_{\nu\sigma}\eta_{\beta\gamma}) 
+2P_3(k_{1\nu}k_{1\gamma}\eta_{\mu\alpha}\eta_{\beta\sigma})
-P_3(k_{1\beta}k_{2\mu}\eta_{\alpha\nu}\eta_{\sigma\gamma}) \nn\\
&&\null \hskip 1.8cm 
 +P_3(k_{1\sigma}k_{2\gamma}\eta_{\mu\nu}\eta_{\alpha\beta})
+P_6(k_{1\sigma}k_{1\gamma}\eta_{\mu\nu}\eta_{\alpha\beta})
+2P_6(k_{1\nu}k_{2\gamma}\eta_{\beta\mu}\eta_{\alpha\sigma}) \nn \\
&&\null \hskip 1.8cm 
+2P_3(k_{1\nu}k_{2\mu}\eta_{\beta\sigma}\eta_{\gamma\alpha})
-2P_3(k_1\cdot k_2 \eta_{\alpha\nu}\eta_{\beta\sigma}\eta_{\gamma\mu})\Bigr]
\,,
\label{deDonderVertex}
\end{eqnarray}
where ``Sym'' signifies a symmetrization in
$\mu\leftrightarrow\alpha$, $\nu\leftrightarrow\beta$ and
$\sigma\leftrightarrow\gamma$, and $P_3$ and $P_6$ signify a
symmetrization over the three external legs, generating three or six
terms respectively.  Here the coupling $\kappa$ is related to Newton's
constant by $\kappa^2 = 32 \pi^2 G_N$. In total the vertex has on the
order of 100 terms. The $k_i$ are the momenta of the three gravitons
and $\eta_{\mu \nu}$ the flat metric.  The precise form of the vertex
depends on the gauge, but in general the three vertex is rather
complicated.  We may contrast this to the relatively simple
three-gluon vertex in Feynman gauge,
\begin{equation}
V^{abc}_{3\mu,\nu,\sigma}(k_1,k_2,k_3) =
g f^{abc} \Bigl[ (k_1 - k_2)_\sigma \eta_{\mu\nu} + \hbox{cyclic} \Bigr] \,.
\label{FeynmanVertex}
\end{equation}
Obviously, based on these considerations gravity would seem much more
complicated than gauge theory.

One can do better with special gauge choices and appropriate field
redefinitions~\cite{vandeVen,Grant}, considerably simplifying
the Feynman rules.  Still, multiloop calculations in gravity based
directly on Feynman diagrams are extremely difficult, if not
impossible, even with the power of modern computers.  Because of this
difficulty, few explicit determinations of counterterms have been
carried out using Feynman diagrams, with some notable
exceptions~\cite{tHooftVeltmanGravity,DeserMatter,GoroffSagnotti,vandeVen}.

\subsection{Power counting in gravity theories}

\begin{figure}[t]
\centerline{\epsfxsize 1.4 truein \epsfbox{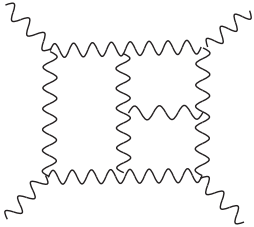}}
\caption[a]{\small A sample diagram at three loops.  A three-graviton
vertex is quadratic in the momenta while a gauge theory vertex in
linear in the momenta, suggesting that gravity theories 
are worse behaved in the ultraviolet than gauge theories. }
\label{ThreeLoopFigure}
\end{figure}

In four dimensions Newton's constant is dimensionful, implying that
gravity theories in four dimensions are non-renormalizable by power
counting. The standard argument that gravity theories are badly
behaved in the ultraviolet follows simply from loop-level Feynman
diagrams.  Consider a Feynman diagram such as the three-loop one in
\fig{ThreeLoopFigure}.  If the diagram is for gluon scattering of
Yang-Mills theory, the diagram yields a Feynman integral of
the form,
\begin{equation}
\int \prod_{j=1}^L \frac{d^D p_j}{ (2 \pi)^D}\;  
   \frac{\cdots (g f^{abc} p_i^\mu) \cdots}{\prod_m (l_m^2 + i \eps)} \,,
\end{equation}
where the numerator factor of the form $(g f^{abc} p_i^\mu)$ signifies
a vertex factor, given by a coupling $g$, color factor $f^{abc}$ and a
momentum $p_j^\mu$. The denominators are the Feynman propagators of
the diagram, carrying momenta depending on the independent loop and
external momenta $p_i$ and $k_i$.  This may be contrasted with the
corresponding gravity diagram, whose form is similar except the
vertices have two powers of momenta in the numerator for each vertex,
\begin{equation}
\int \prod_{j=1}^L \frac{d^D p_j}{(2 \pi)^D} \;
 \frac{\cdots (\kappa p_i^\mu p_k^\nu) \cdots }{\prod_m (l_m^2 + i \eps)} \,.
\end{equation}
Generically, the momenta appearing in the vertices will be loop
momenta.  Because of the larger number of loop momenta in the
numerators we may expect each gravity Feynman integral to be worse
behaved in the ultraviolet, compared to the corresponding gauge theory
diagram, unless there are non-trivial cancellations.  Based on these
simple power-counting considerations, as the number of loops increases
we may expect an ever-worsening behavior compared to gauge theory.

We can easily determine the structure of a counterterm given a
divergence at a given loop order.  Since every
loop gains an extra power of $G_N \sim 1/M_{\rm planck}^2$ every
additional loop must gain two powers of mass dimension to
compensate. For graviton amplitudes this corresponds to an additional
power of the Riemann tensor, $R_{\mu \nu\sigma\rho}$, or to a
covariant derivative ${\cal D}^2$.  At one loop, pure-gravity
counterterms in the Lagrangian involve two powers of the Riemann
tensor, but by an accident of four dimensions, the Gauss-Bonnet
theorem eliminates the potential on-shell counterterm.
However, if matter is added, then generically a one-loop divergence
appears~\cite{tHooftVeltmanGravity,DeserMatter}.  At two loops the
potential counterterm is of the form $ R^{\lambda \rho}{}_{\mu\nu}
R^{\mu\nu}{}_{\sigma \tau} R^{\sigma \tau}{}_{\lambda\rho}$. In pure
gravity, thanks to calculations by Goroff and Sagnotti and by
van~de~Ven~\cite{GoroffSagnotti,vandeVen}, we know for a fact that the theory
diverges and the coefficient of this counterterm is nonvanishing.

The first divergence in any four-dimension supergravity theory can be
no earlier than three loops, since the potential one- and two-loop
counterterms cannot be made consistent with
supersymmetry~\cite{GrisaruTomboulis}.  In this case, the potential
counterterm consistent with supersymmetry is an $R^4$
term~\cite{Supergravity} with indices appropriately contracted,
corresponding to the square of the Bel-Robinson tensor.  With
additional assumptions we can raise the loop order where a divergence
is first expected in $\NeqEight$ supergravity.  For example, if an
off-shell superspace with ${\cal N} = 6$ supersymmetries manifest were
to exist, potential divergences in four dimensions would be delayed to
at least five loops~\cite{HoweStelleNew}.  This is supported by recent
supersymmetry arguments which do not rely on the existence of such a
superspace~\cite{HoweStelleRecent}.  One can go beyond this by
assuming the existence of a superspace with ${\cal N}=7$
supersymmetries manifest which would delay the first potential
divergence to at least six loops~\cite{HoweStelleNew}. Continuing in
this way, if one were to assume the existence of a fully covariant
off-shell superspace with ${\cal N}=8$ supersymmetries manifest, then
the first potential divergence would be pushed to seven
loops~\cite{GrisaruSiegel}.  It is important to note that since no
off-shell superspace beyond ${\cal N}=4$ has been constructed for
supergravity theories, bounds based on assuming their existence are
not firm.  A full superspace invariant, which could act as potential a
counterterm, has been constructed at eight loops, suggesting that a
divergence might appear at this loop order, if it does not appear
earlier~\cite{Kallosh}.  The appearance of the first potential
divergence can even be pushed to nine loops, with an additional
speculative assumption that all fields respect ten-dimensional general
coordinate invariance~\cite{KellyPrivate}.  This bound coincides with
the one argued~\cite{GreenII} from the type~II string theory
non-renormalization theorem of Berkovits~\cite{Berkovits}.  Beyond
this order, no purely supersymmetric mechanism has been suggested for
preventing the onset of divergences.  Indeed, for a supergravity
theory to be ultraviolet finite to all orders of the perturbative
expansion, novel cancellations beyond the known supersymmetric ones
{\it must} exist.

Of course, no power counting arguments can {\it prove} the existence
of a divergence, only that divergences cannot appear prior to a
certain loop order.  If the theory were to possess a hidden symmetry,
not accounted for in the power counting bound, the bound may suggest
the appearance of a divergence when, in fact, there is none.  The
possibility of hidden symmetries is real: the three-loop divergence
thought to appear in the three-loop four-point amplitude in the
studies from 1980's~\cite{Supergravity} is now known to not be
present~\cite{GravityThree, CompactThree,HoweStelleRecent}. As already
mentioned in the introduction, there are a number of good reasons to
question the conventional wisdom and to reexamine the ultraviolet
behavior of $\NeqEight$ supergravity using modern on-shell methods.

\section{On-shell methods}
\label{OnShellSection}

The computations we summarize in the next section were obtained using
on-shell methods. A key feature of these methods is that the
elementary building blocks for obtaining new amplitudes are previously
obtained on-shell amplitudes.  These methods fall into two basic
categories: on-shell recursion~\cite{BCFRecursion,BCFW} and the modern
unitarity method~\cite{UnitarityMethod}.  On-shell recursion is
suitable for obtaining complete tree-level amplitudes while the
unitarity method is suitable at loop level.  At one loop a hybrid
bootstrap approach making use of both unitarity and on-shell recursion
has also been developed~\cite{Bootstrap}.  The use of these methods
have become much more widespread in recent years because of their
computational efficiency, especially at loop level.  These methods
have been applied to a wide variety of problems, including QCD (see
refs.~\cite{QCDApplications} for examples of recent applications),
AdS/CFT studies of $\NeqFour$ super-Yang-Mills
amplitudes~\cite{Iterate,SixPointTwoLoop}, and to quantum gravity which
we discuss here.  Here we only briefly review on-shell methods, focusing
on the salient features for gravity, and refer the reader to various
reviews~\cite{Review} for more extensive expositions.

\subsection{On-shell vertices}

As an explicit example of the inherent simplification of on-shell
methods, consider an on-shell version of the three-graviton vertex in
\eqn{deDonderVertex}, where we dot the three legs with physical
polarizations tensors, satisfying the on-shell conditions,
$k_i^2 = \pol_i^{\mu \nu}  k_{i\mu} =\pol_i^{\mu \nu} k_{i\nu} =
\pol^\mu{}_\mu = 0$.  This gives the simplified vertex,
\begin{eqnarray}
&& G_{3}(k_1,k_2,k_3)
 =  - i \kappa \pol_1^{\mu \alpha} 
         \pol_2^{\nu \beta}    \pol_3^{\sigma \gamma}
  \Bigl[ (k_1)_\sigma \eta_{\mu\nu} + \hbox{cyclic} \Bigr]
           \Bigl[ (k_1)_\gamma \eta_{\alpha\beta} + \hbox{cyclic} \Bigr] \,,
\label{OnShellVertex}
\end{eqnarray}
which is not much more complicated than the corresponding 
on-shell Yang-Mills vertex,
\begin{equation}
V_{3}^{abc}(k_1,k_2,k_3) = \pol_1^{\mu} 
         \pol_2^{\nu}    \pol_3^{\sigma}
2 g f^{abc} \Bigl[ (k_1)_\sigma \eta_{\mu\nu} + \hbox{cyclic} \Bigr] \,,
\label{OnShellYMVertex}
\end{equation}
where the polarization vector satisfies $\pol_i^{\mu} k_{i\mu} = 0$.
As we shall discuss below, the structure of the on-shell graviton
vertex as a product of the kinematic part of gauge-theory vertices is
{\it not} accidental, but reflects a profound and important property of
gravity.

If we proceed naively with an on-shell formalism, we encounter a
difficulty: the on-shell three vertex for gravitons or gluons actually
vanishes, because the process is kinematically forbidden.  The
solution is to use complex
momenta~\cite{GoroffSagnotti,ComplexMomenta,BCFUnitarity}, which allows us to
satisfy the on-shell conditions and momentum conservation, yet define
a nonvanishing vertex. 

In general, in four dimensions it is best to express gauge and
gravity amplitudes in terms of spinors.  We define the two helicity
configurations gluons as
\begin{equation}
\pol_\pm^\mu(k_i, q_i) = \pm \frac{\sand{q_i}.{\gamma^\mu}.{k_i}}{
                  \sqrt{2} \langle q_i^\mp | k_i^\pm \rangle } \,,
\end{equation}
where $|k_i^\pm \rangle \equiv  u_\pm(k_l)$ are Weyl spinors and   
the $q_i$ are arbitrary null ``reference momenta''.
The graviton polarization tensors are simply products of these,
\begin{equation}
\pol_\pm^{\mu\alpha}(k_i, q_i) = 
 \pol_\pm^{\mu}(k_i, q_i) \, \pol_\pm^{\alpha}(k_i, q_i)  \,.
\end{equation}
For later reference we define spinor inner products, as
\begin{equation}
\spa{j}.{l} = \langle{k_j^-}|{k_l^+}\rangle \,, \hskip 3 cm 
\spb{j}.{l} = \langle{k_j^+}|{k_l^-}\rangle \,.
\end{equation}
These products are antisymmetric, $\spa{j}.{l} = - \spa{l}.{j}$,
$\spb{j}.{l} = - \spb{l}.{j}$.  For massless momenta, spinor inner
products are complex square roots of Lorentz inner products,
satisfying $\spa{i}.{j} \spb{j}.{i} = 2 k_i \cdot k_j$.  For further
details, see refs.~\cite{TreeReview}.


\subsection{Behavior of amplitudes under large complex shifts}
\label{OnShellRecursionSubsection}

On-shell recursion~\cite{BCFRecursion, BCFW,GravityRecursion,
AHK} gives us a simple means for systematically
constructing new tree-level amplitudes in gravity and in gauge theory
using previously constructed lower-point amplitudes.  The proof of
the recursion relations employs a complex shift of two momenta,
\begin{eqnarray}
&k_j^\mu &\rightarrow k_j^\mu(z) = k_j^\mu -
       \frac{z}{2}{\sand{j}.{\gamma^\mu}.{l}},\nonumber\\
&k_l^\mu &\rightarrow k_l^\mu(z) = k_l^\mu +
       \frac{z}{2}{\sand{j}.{\gamma^\mu}.{l}} \,,
\label{MomentumShift}
\end{eqnarray}
so that they remain massless, $k_j^2(z) = k_l^2(z) = 0$, and overall
momentum conservation is maintained.   An on-shell
amplitude containing the momenta $k_j$ and $k_l$ then
depends on the parameter $z$,
\begin{equation}
A(z) = A(k_1,\ldots,k_j(z),k_{j+1},\ldots,k_l(z),k_{l+1},\ldots,k_n)\,,
\end{equation}
where the physical amplitude is recovered by setting $z = 0$.

A crucial step in the proof of on-shell recursion relations is the
requirement that $A(z)$ vanishes as $z\rightarrow\infty$.  This
property has been demonstrated for a wide class of shifts in
gauge~\cite{BCFW, GloverMassive} and gravity
theories~\cite{GravityRecursion,AHK}.  This property is especially
surprising for gravity, where the larger number of powers of loop
momentum in the vertices would tend to make the diagrams ill-behaved
as $z \rightarrow \infty$. In general, individual Feynman diagrams are
not well behaved as $z\rightarrow\infty$; only the complete amplitudes
vanish in this limit because of a strong cancellation between
diagrams.  However, with a space-like gauge similar to light-cone
gauge~\cite{SpaceCone} it is possible to arrange the diagrams in both gauge
theory~\cite{VamanYao} and gravity~\cite{AHK} such that each diagram
is well behaved at large $z$.  The improved large $z$ behavior appears
connected to an enhanced Lorentz symmetry appearing in this limit~\cite{AHK}.

An important consequence of the on-shell recursion relations is that
we can recursively construct tree-level scattering amplitudes starting
from on-shell three-point vertices. This can even be carried out for
pure gravity in $D$ dimensions~\cite{AHK}.  Using the unitarity method
these tree amplitudes are all that are needed to systematically
construct all higher-loop amplitudes.  This construction does not use
four- and higher-point vertices in any step for either gauge or
gravity theories.  Thus, rather surprisingly the four- and
higher-point vertices displayed in \figs{YMVertFigure}{GravVertFigure}
are irrelevant, as far as scattering amplitudes to any loop order are
concerned.

Note the behavior of the amplitudes as $z\rightarrow \infty$ is
connected to the high-energy behavior of the theory, albeit in a
complex direction.  As we shall discuss below, the vanishing of $A(z)$
as $z \rightarrow \infty$ is connected directly to loop-level ultraviolet
cancellations~\cite{NoTri}.

\subsection{Kawai-Lewellen-Tye tree-level relations}

From the perspective of Lagrangians or off-shell Feynman rules it is very
difficult to discern any simple relations between gravity and gauge
theory amplitudes.  Nevertheless, tree-level gravity amplitudes can be
rewritten in terms of gauge-theory amplitudes, a fact first uncovered
in string theory by Kawai, Lewellen and Tye (KLT)~\cite{KLT, Grant,
FreitasGravity}.  These relations also hold in field
theory, as the low-energy limit of string theory.  In this limit, the
KLT relations for four-, and five-point amplitudes are
\begin{eqnarray}
M_4^\tree (1,2,3,4) &\! =\!& - i s_{12} A_4^\tree(1,2,3,4) \,
   \widetilde{A}_4^\tree(1,2,4,3)\,, \label{KLT4}\\
M_5^\tree(1,2,3,4,5) 
&\! = \!& i s_{12} s_{34}  A_5^\tree(1,2,3,4,5) \,
                \widetilde{A}_5^\tree(2,1,4,3,5) \nn \\
&& \hskip 2 cm \null 
+ i s_{13}s_{24} A_5^\tree(1,3,2,4,5) \, 
           \widetilde{A}_5^\tree(3,1,4,2,5) \,.\label{KLT5}
\end{eqnarray}
Here the $M_n$'s are amplitudes in a gravity theory stripped of
couplings, the $A_n$'s and $\widetilde{A}_n$'s are two distinct
color-ordered gauge theory amplitudes and the Mandelstam invariants
are $s_{ij} \equiv (k_i+k_j)^2$, with $k_i$ being the outgoing momentum
of leg $i$.  The color-ordered~\cite{TreeReview} gauge-theory
amplitudes correspond to the coefficient of color traces with a given
ordering of matrices. The gravity states are direct products of
gauge-theory states for each external leg. Explicit formul\ae{} for
$n$-point amplitudes may be found in
refs.~\cite{OneloopMHVGravity,Twist}.

As a simple example, consider the four-graviton amplitude
$M_4^\tree(1^-,2^-, 3^+,4^+)$, where the $\pm$ labels refer to the
helicity of the gravitons in an all outgoing convention.  Using the
four-gluon color-ordered amplitudes~\cite{TreeReview},
\begin{eqnarray}
A_4^\tree(1^-,2^-, 3^+,4^+) = 
i  \frac{\spa1.2^3}{\spa2.3 \spa3.4 \spa4.1} \,, \nn\\ 
A_4^\tree(1^-,2^-, 4^+,3^+) = 
i  \frac{\spa1.2^3}{\spa2.4 \spa4.3 \spa3.1} \,, 
\end{eqnarray}
from the KLT relation (\ref{KLT4}), we obtain immediately,
\begin{equation}
{\cal M}_4^\tree(1^-,2^-, 3^+,4^+) = i\biggl( \frac{\kappa}{2} \biggr)^2
s_{12}\, \frac{\spa1.2^3}{\spa2.3 \spa3.4 \spa4.1}  \,
       \frac{\spa1.2^3}{\spa2.4 \spa4.3 \spa3.1} \,,
\label{FourGravitonExample}
\end{equation}
where we have restored the gravity coupling constant.  We invite the
reader to evaluate this four-graviton amplitude using Feynman rules
starting from the Einstein action: the result will match
\eqn{FourGravitonExample}.  These relations hold not only for any
theory corresponding to the low energy limit of a string theory, but
appear to hold for a much broader range of gravity
theories~\cite{FreitasGravity}.

The KLT relations have recently been clarified, giving a more
transparent form of the relation~\cite{Twist}.  Consider the
diagrammatic expansion of color-dressed gauge-theory amplitudes,
\begin{eqnarray}
&& {\cal A}^\tree_n(1,2,3,\ldots,n)= 
  g^{n-2}\sum_{i}
                   \frac{n_i c_i }{(\prod_{j} p^2_j)_i}\, , \nn \\
&& {\cal \widetilde{A}}^\tree_n(1,2,3,\ldots,n)= \tilde g^{n-2} \sum_{i}
                   \frac{\tilde n_i c_i}{(\prod_{j} p^2_j)_i} \, ,
\label{GaugeTheoryLeftRight}
\end{eqnarray}
where the ${\cal A}^\tree_n$ and ${\cal \widetilde{A}}^\tree_n$ are
two distinct gauge-theory amplitudes including color factors, $c_i$,
and $g$ is the coupling constant. The sum runs over all diagrams
with only cubic vertices, so each diagram has exactly $n-3$
propagators. The denominator ${(\prod_{j} p^2_j)_i}$, represents the
Feynman propagators of the $i^{\rm th}$ diagram.  As discussed in
ref.~\cite{Twist} the $n_i$ and $\tilde n_i$ numerators can be
constrained to satisfy a kinematic numerator identity similar to the
Jacobi identity satisfied by the $c_i$ color factors.  If the
numerators satisfy this constraint, the gravity amplitudes, including
factors of the coupling $\kappa$, are then given by,
\begin{eqnarray}
&&  {\cal M}^\tree_n(1,2,3,\ldots,n)= i \biggl(\frac{\kappa}{2}\biggr)^{n-2}
              \sum_{i} \frac{n_i \tilde n_i}{(\prod_{j} p^2_j)_i} \, .
\label{GravityMasterFormula}
\end{eqnarray}
The sum runs over the same set of diagrams as for the gauge theory case
in \eqn{GaugeTheoryLeftRight}. This equation has been explicitly
checked to be equivalent to the KLT relations 
through eight points~\cite{Twist}.

Although the KLT relations are at present used merely as a technical
trick to efficiently evaluate gravity amplitudes, they point towards a
non-trivial field-theory unification of gravity and gauge theory.  As
is well known, string theory automatically encodes this unification.

\subsection{Modern unitarity method for obtaining loop amplitudes}

\begin{figure}[t]
\centerline{\epsfxsize 6.0 truein \epsfbox{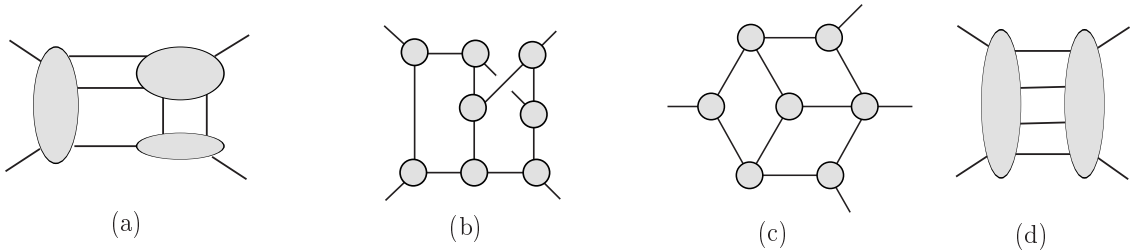}}
\caption[a]{\small Examples of generalized cuts for determining a
three-loop four-point amplitude. Each blob represents an on-shell tree
amplitude, and the displayed intermediate lines are all on shell.
}
\label{SampleCutsFigure}
\end{figure}

The modern unitarity method~\cite{UnitarityMethod} gives us a
systematic means for constructing loop amplitudes using on-shell tree
amplitudes as input.  To obtain an amplitude one first constructs an
initial ansatz in terms of integral functions that reproduces one cut.
Then, subsequent cuts of the amplitude are compared against the
corresponding cuts of the ansatz. If any discrepancy is found in a
later cut, additional terms that vanish when all the earlier cut
conditions are imposed are added to the ansatz.  Once a complete set
of cuts have been checked, this procedure gives an integral
representation of the loop amplitude with the correct cuts in all
channels.  The result is the complete amplitude, equivalent to what
would have been obtained with Feynman diagrams.

At one loop, massless amplitudes in supersymmetric gauge theories are
determined completely by their four-dimensional
cuts~\cite{UnitarityMethod}.  Unfortunately, this has not been
demonstrated at higher loops. Even if the theory is ultraviolet
finite, infrared singularities are present in four dimensions
requiring use of a version of dimensional regularization compatible
with supersymmetry~\cite{DimRed}.  To guarantee that no terms are
dropped in the construction, the unitarity cuts must be evaluated in
$D$ dimensions.  Unfortunately, evaluating the cuts in $D$
dimensions~\cite{DDimUnitarity} makes the calculation significantly
more difficult, because powerful four-dimensional spinor
methods~\cite{TreeReview} can no longer be used.\footnote{An
interesting paper with a six-dimension helicity-like formalism
recently appeared~\cite{D6Helicity}.}  For $\NeqFour$ super-Yang-Mills
theory, some of this additional complexity can be sidestepped by
performing internal-state sums in terms of the gauge supermultiplet of
$D=10,\, {\cal N}=1$ super-Yang-Mills theory instead of the $D=4,\,
\NeqFour$ multiplet; this counts the same states but simplifies the
bookkeeping. 

In practical calculations it is useful to first construct an ansatz
for the amplitude based on using four-dimensional momenta in the
generalized cuts, since we can use powerful spinor methods. Recently,
there has been considerable improvement in bookkeeping the
superpartners crossing unitarity cuts in four
dimensions~\cite{RecentOnShellSuperSpace,AHCKGravity}, based on
extensions of Nair's on-shell superspace~\cite{Nair} to general
amplitudes. After an ansatz for the complete amplitudes is constructed
based on four-dimensional cuts, to guarantee that no terms are
dropped, it must be check against the $D$-dimensional cuts.  However,
for four-point amplitudes in $\NeqFour$ super-Yang-Mills theory the
weight of evidence points to all terms being detectable through five
loops by four-dimensional cuts~\cite{BRY, BDDPR, FourLoop, FiveLoop}.
Beyond this we do not expect cuts with four dimensional kinematics to
be sufficient.  Indeed, for two-loop six-point amplitudes, terms
appear that vanish in four dimensions~\cite{SixPointTwoLoop}.

The KLT relations give us a rather efficient means of evaluating
gravity generalized cuts that reduce the amplitude to products of tree
amplitudes summed over intermediate states.  These relations allow us
to re-express cuts of gravity amplitudes to sums of products of cuts
of gauge-theory amplitudes.  The gauge-theory cuts are generally much
simpler to evaluate.  An important feature of this construction is
that once the superpartner sums are performed for $\NeqFour$
super-Yang-Mills generalized cuts, the corresponding super-partner sum
in $\NeqEight$ supergravity follows directly from the KLT relations.
This holds in $D$ dimensions as well.  More generally, any
simplifications performed on the gauge-theory generalized cuts
carry over immediately to gravity cuts.

\section{Ultraviolet properties of $\NeqEight$ supergravity}
\label{ReviewSection}

\subsection{Early higher-loop computations}

In refs.~\cite{BRY,BDDPR}, the two-loop four-point amplitudes of
$\NeqFour$ super-Yang-Mills theory and $\NeqEight$ supergravity were
determined in terms of scalar integrals via the unitarity method.  A
class of unitarity cuts---the ``iterated two-particle cuts''---were
evaluated to all loop orders, yielding a partial reconstruction of the
four-point amplitudes at any loop order. The iterated cuts are fairly
simple to evaluate because the same sewing algebra appears at each
loop order~\cite{BRY}.  For the case of $\NeqFour$ super-Yang-Mills
theory, examining the powers of loop momenta in the numerator of the
generic iterated two-particle cut contributions suggests the
finiteness bound~\cite{BDDPR},
\begin{equation}
D < \frac{6}{L} + 4  \hskip 1 cm (L > 1) \,,
\label{SuperYangMillsPowerCount}
\end{equation}
where $D$ is the dimension of space-time and $L$ the loop order.  (The
case of one loop, $L=1$, is special, with the amplitudes finite for
$D<8$, not $D<10$.)  The bound~(\ref{SuperYangMillsPowerCount})
differs somewhat from earlier superspace power
counting~\cite{HoweStelleYangMills}, although all bounds confirm the
ultraviolet finiteness of $\NeqFour$ super-Yang-Mills theory in
$D=4$. This all-loop-order bound (\ref{SuperYangMillsPowerCount}) has
since been confirmed~\cite{HoweStelleNew} using ${\cal N}= \nobreak 3$
harmonic superspace~\cite{HarmonicSuperspace}. Explicit computations
demonstrate this bound is saturated through at least four
loops~\cite{BRY,BDDPR,Finite}.

In ref.~\cite{BDDPR}, the iterated two-particle cuts of $\NeqEight$
supergravity amplitudes were also analyzed, leading to the proposal that
the four-point $\NeqEight$ supergravity amplitude should be ultraviolet finite
for
\begin{equation}
D< \frac{10}{L} + 2  \hskip 1 cm  (L>1)\,.
\label{OldPowerCount}
\end{equation}
(Again the one-loop case is special with the $\NeqEight$ supergravity
amplitudes being finite for $D<8$~\cite{NoTriangleProof}.)  This bound
implies that in $D=4$ the first potential divergence in $\NeqEight$
supergravity may appear at five loops.  The formula is also consistent
with bounds obtained by Howe and Stelle~\cite{HoweStelleNew}, assuming
the existence of an ${\cal N} = 6$ harmonic
superspace~\cite{HarmonicSuperspace}. It is also consistent with a
newer analysis~\cite{HoweStelleRecent} predicting divergences for
$L=4,D=5$ and $L=5,D=4$, but which does not rely on the existence of
such a superspace.  Since bound (\ref{OldPowerCount}), proposed in
ref.~\cite{BDDPR} is based on a partial calculation, if there are
additional hidden cancellations with uncalculated terms then the true
finiteness bound would be improved compared to \eqn{OldPowerCount}.

\subsection{Novel cancellations at one loop}

\begin{figure}[t]
\centerline{\epsfxsize 4 truein \epsfbox{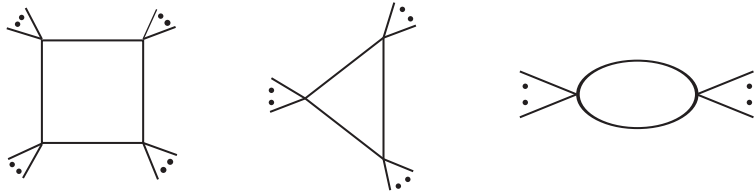}}
\caption[a]{\small Any one-loop $n$-point amplitude can be expressed
as a linear combination of scalar box, triangle and bubble integrals.
In $\NeqFour$ super-Yang-Mills and $\NeqEight$ supergravity, after
reducing all integrals to this basis, the coefficients of all triangle
and bubble integrals vanish.}
\label{OneloopBasisFigure}
\end{figure}

A key clue for the existence of novel cancellations at higher
loops~\cite{Finite} comes from one loop.  A one-loop
theorem~\cite{IntegralBasis} states that near four dimensions any
$n$-point massless amplitude can be expressed as
\begin{equation}
A_n = \sum_i d_i I_4^i + \sum_i c_i I_3^i + \sum_i b_i I_2^i \,, 
\label{IntegralBasis}
\end{equation}
where the scalar integrals $I_{2,3,4}$ are respectively bubbles,
triangles, and boxes and $d_i,c_i, b_i$ are (possibly dimension
dependent) rational coefficients.  These basis scalar integrals are
displayed in \fig{OneloopBasisFigure} and are the same ones that would
appear in $\varphi^3$ theory.  Some time ago, for the special case of
maximally helicity violating one-loop amplitudes of $\NeqEight$
supergravity, a curious cancellation was observed, setting the
coefficients of all triangle and bubble integrals to
zero~\cite{OneloopMHVGravity}.  More recently, it became clear that
the vanishing of the triangle and bubble integral coefficient was a
general property of all one-loop $\NeqEight$ supergravity
amplitudes~\cite{NoTriangle, NoTriangleB, BjerrumVanhove}, with proofs given in
refs.~\cite{NoTriangleProof,AHK}.  This has been called the ``no-triangle
property'', although it also implies lack of bubble integrals.  For
$\NeqEight$ supergravity this one-loop behavior is rather surprising
and suggests the existence of hidden cancellations not apparent in
naive power counting.  We can use this to limit the number of powers
of loop momenta which can appear in the numerator of one-loop
amplitudes in $\NeqEight$ supergravity.  Under the Brown-Feynman or
Passarino-Veltman reduction~\cite{IntegralBasis} each power of $l
\cdot k_i$---where $l$ is a loop momentum and $k_i$ an external
momentum---will give a linear combination of integrals with either no
or one canceled propagator.  For example, an amplitude with a hexagon
integral with a numerator of the form $(l \cdot k_i)^3$, after
integral reductions will contain box and triangle integrals.  With an
additional power of $l \cdot k_i$, bubble integrals will also appear.

Interestingly, the observed novel cancellations appear to be generic
in gravity theories, as suggested by the one-loop study of
ref.~\cite{NoTri}.  In fact, these type of cancellations may be
generic to any theory where color factors do not prevent cancellations
between contributions with a differing ordering of
legs~\cite{UnorderedBjerrum}.  The novel one-loop cancellations were
also shown to follow from the remarkably good high-energy behavior of
gravity tree amplitudes under the complex deformations used to prove
on-shell recursion relations in gravity~\cite{BCFW,
GravityRecursion, AHCKGravity}.

\begin{figure}[t]
\centerline{\epsfxsize 1.5 truein \epsfbox{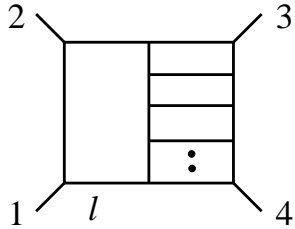}}
\caption[a]{\small An $L$-loop contribution as proposed in
ref.~\cite{BDDPR} based on a partial analysis of unitarity cuts. In
$\NeqFour$ super-Yang-Mills theory the numerator factor is $[(l+
\nobreak k_4)^2]^{(L-2)}$, after dividing out factors which are
independent of loop momenta. In the $\NeqEight$ supergravity case the
proposed corresponding factor was $[(l+k_4)^2]^{2(L-2)}$.  This
proposal violates the no-triangle property of the one-loop amplitude
isolated by the cut in \fig{NCutFigure}(a), implying additional 
cancellations with other contributions. }
\label{NOldPowerFigure}
\end{figure}

\begin{figure}[t]
\centerline{\epsfxsize 3.5 truein \epsfbox{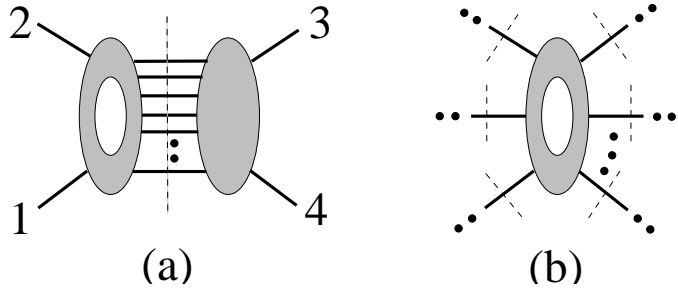}}
\caption[a]{\small The no-triangle property ensures that all one-loop
subamplitudes appearing in a multiloop $\NeqEight$ supergravity
amplitude have the same degree of divergence as that of $\NeqFour$
super-Yang-Mills theory.  The cut (a) is an $L$-particle cut of an
$L$-loop amplitude. Cut (b) makes use of generalized unitarity to
isolate a one-loop subamplitude.  These cuts may be used to
demonstrate the existence of novel cancellations to all loop orders.}
\label{NCutFigure}
\end{figure}

\subsection{Novel cancellations to all loop orders}

By itself, the cancellation of one-loop triangle and bubble integrals
at $n$-points does not directly say anything about the ultraviolet
properties of the theory. However, as pointed out in
ref.~\cite{Finite}, for a certain class of terms, the unitarity method
gives us a powerful means to non-trivially constrain the ultraviolet
behavior of multiloop amplitudes to {\it all} loop
orders~\cite{Finite}, starting from the one-loop no-triangle
cancellations.  In particular, consider the diagrammatic contributions
to the $L$-loop four-point amplitude of the form in
\fig{NOldPowerFigure}.  Based on evaluating the iterated two-particle
cuts, ref.~\cite{BDDPR} suggested that for $\NeqFour$ super-Yang-Mills
theory the numerator of the integral corresponding to this diagram be
dressed by a factor of
\begin{equation}
[(l +k_4)^2]^{(L-2)}\,,
\label{N4Numerator}
\end{equation}
after removing factors which are independent of the loop momentum $l$.  For
$\NeqEight$ supergravity, the proposed corresponding factor was,
\begin{equation}
[(l+k_4)^2]^{2(L-2)}\,.
\label{N8Numerator}
\end{equation}
which is much worse behaved as $L$ increases compared to the
super-Yang-Mills numerator (\ref{N4Numerator}).  Power counting the
integral in \fig{NCutFigure} with this numerator gives the $\NeqEight$
finiteness bound in \eqn{OldPowerCount}.

However, additional cancellations must exist because the too high a
power of loop momentum in the proposed numerator (\ref{N8Numerator})
would lead to a violation of the no-triangle-integral property in the
one-loop subamplitude isolated by the unitarity cut shown in
\fig{NCutFigure}(a). This violation would start at three loops.  To
prevent this, non-trivial cancellations with other integrals would
need to start at three loops, effectively reducing the number of
powers of loop momenta in the numerators.  Besides constraints on the
power counting from the cut in \fig{NCutFigure}(a), non-trivial
additional constraints follow from the set of all generalized cuts
isolating a one-loop subamplitude, schematically depicted in
\fig{NCutFigure}(b).  After the cancellations take place, at three
loops only two powers of loop momenta should remain, matching the
number appearing in the $\NeqFour$ super-Yang-Mills case
(\ref{N4Numerator}).  This has been confirmed by explicit
calculation~\cite{GravityThree, CompactThree}.

It is interesting that there does not appear to be a purely
supersymmetric explanation of these cancellations~\cite{HoweStelleNew,
HoweStelleRecent}.  If it is not entirely supersymmetry then what
might be behind the cancellations?  The unitarity method implies that
cancellations must have their origin in tree-level cancellations.  An
obvious guess is that the loop-level cancellations should be related
to the tree-level ones under large complex deformations of the momenta
discussed in \sect{OnShellRecursionSubsection}. Indeed this is the
case~\cite{NoTriangleB,NoTri}.  From an analysis of one-loop
cancellations using an integration formalism due to
Forde~\cite{Forde}, the following picture emerges~\cite{NoTri}: Most
of the one-loop cancellations observed in $\NeqEight$ supergravity
leading to the no-triangle property are already present in
non-supersymmetric pure gravity and are directly tied to the
tree-level cancellations.  Schematically, in $\NeqEight$ supergravity
we find that a one-loop integral with the $2n$ maximum number of
powers of loop momentum undergoes a cancellations of $n+4$ powers,
leaving $n-4$ powers.  Of the canceled powers of loop momentum, $8$
are due to supersymmetry while $n-4$ are generic and cancel also in
pure Einstein gravity.  The generic cancellations have recently been
attributed to the unordered nature of the graviton
amplitudes~\cite{NoTriangleProof, UnorderedBjerrum}.

Although the above arguments demonstrate the existence of novel
ultraviolet cancellations to {\it all} loop orders, they do not
demonstrate that these cancellations exist for all contribution.  In
particular, contributions not detectable by isolating a one-loop
subamplitude will not be constrained.  We therefore need to inspect
such terms. The three-loop four-point amplitude is the logical place
to do so because it is the simplest example expected to exhibit the
novel cancellations.

\subsection{Superfiniteness at three loops}

In refs.~\cite{GravityThree,CompactThree} the full three-loop
four-point amplitude of $\NeqEight$ supergravity was obtained using
the unitarity method.  These computations establish the existence of
novel cancellations in terms not sensitive to cuts isolating a
one-loop subamplitude.

In the original calculation of the three-loop four-point
amplitude~\cite{GravityThree}, the result was given in a form where
individual diagrams have a worse power counting than the complete
answer. An improved form, was subsequently found using the method of
maximal cuts~\cite{FiveLoop}.\footnote{A related method is the leading
singularity method, which has also been used in calculations of
various maximally supersymmetric multiloop
amplitudes~\cite{LeadingSingularity}.}  In the latter form the
individual contributions are no worse behaved than the contributions
of the corresponding super-Yang-Mills amplitude~\cite{CompactThree}.

\begin{figure}[t]
\centerline{\epsfxsize 5.5 truein \epsfbox{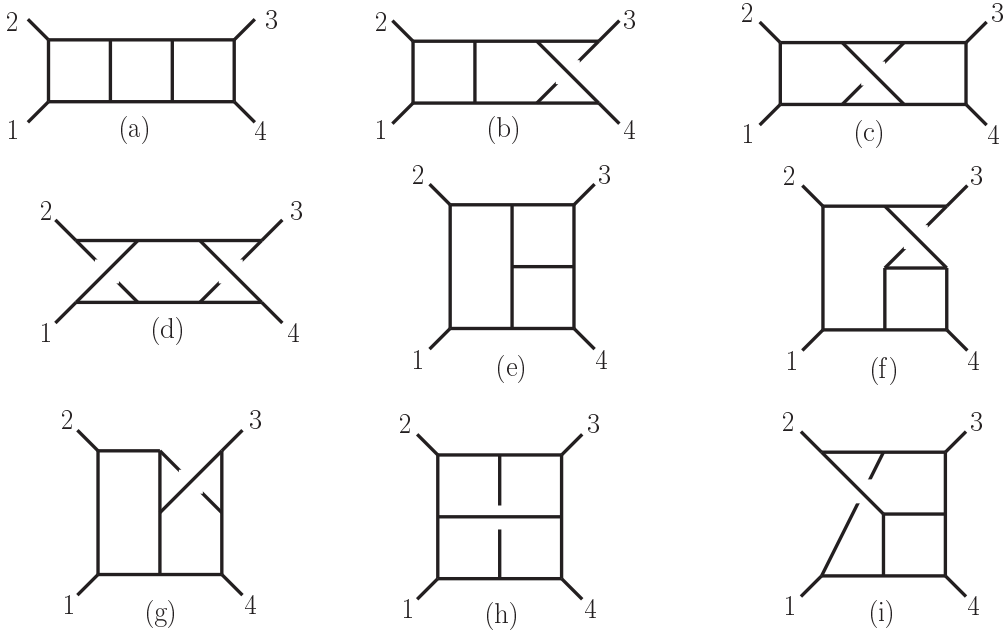}}
\caption[a]{\small The different parent diagrams in terms of which 
four-point three-loop amplitudes may be expressed.}
\label{IntegralsThreeLoopFigure}
\end{figure} 

The three-loop amplitudes can be described in terms of the ``parent
diagrams'' displayed in \fig{IntegralsThreeLoopFigure} dressed by
appropriate numerator factors.  The integrals composing the amplitudes
are of the form,
\begin{equation}
I^{(x)}= i 
\int  \Biggl[ \prod_{i=1}^3 \frac{d^D q_i}{(2\pi)^D} \Biggr] \, 
 \frac{N^{(x)}(q_j, k_j) }{\prod_{m=5}^{14} l_m^2} \,,
\label{IntegralNormalization}
\end{equation}
where the $q_i$'s are three independent loop momenta, the $l_m$'s are
the momenta carried by the propagators of the diagrams, and the
$N^{(x)}(q_j, k_j) $ are numerator factors.

In the improved form given in ref.~\cite{CompactThree}, no  
numerator factors of $\NeqEight$ supergravity have a worse power count 
than the corresponding factors of $\NeqFour$ super-Yang-Mills theory.
That is, numerators in either theory satisfy,
\begin{equation}
\partial_{q_i}\partial_{q_j} \partial_{q_k}  N^{(x)}(q_m,k_m) = 0 \,,
\end{equation}
for the independent loop momenta with $i,j,k$ taking on the values
$1,2,3$.  Thus, no more than two powers of loop momenta appear and the
power counting of the three-loop four-point $\NeqEight$ supergravity
amplitude is identical to that of the corresponding $\NeqFour$
super-Yang-Mills amplitude.  Not only does this computation prove that
the three-loop divergence is not present in four dimensions, the
amplitude exhibits ``superfiniteness'' or cancellations beyond those
required for finiteness.  Instead of the finiteness
condition~(\ref{OldPowerCount}), at $L=3$ the correct power count for
the four-point amplitude is the one matching the corresponding
super-Yang-Mills amplitude (\ref{SuperYangMillsPowerCount}).  Explicit
calculations show that the finiteness bound
(\ref{SuperYangMillsPowerCount}) is saturated in both theories, so at
three loops both theories do diverge in six dimensions.

\section{Conclusions}
\label{ConclusionSection}

Maximally supersymmetric $\Neqeight$ supergravity theory is the most
promising candidate for a unitary perturbatively
ultraviolet-finite point-like quantum field theory of gravity.  The
modern unitarity method offers a powerful means to probe its multiloop
ultraviolet behavior, mitigating the rather rapid
increase in complexity of Feynman diagrams at higher-loop
orders.  We summarized the results of refs.~\cite{GravityThree,
CompactThree}, where a loop-integral representation of the three-loop
four-point amplitude of $\NeqEight$ supergravity was presented,
exhibiting cancellations beyond those needed for finiteness.  The
three-loop power counting matched the one for $\NeqFour$
super-Yang-Mills theory, which is known to be ultraviolet finite.

As we discussed, for a subset of contributions novel
all-loop cancellations exist with no known supersymmetry explanation.
This follows from an investigation of a class of generalized unitarity
cuts~\cite{Finite}, making use of the one-loop ``no-triangle
property''~\cite{OneloopMHVGravity, NoTriangle, NoTriangleB,
NoTriangleKallosh, BjerrumVanhove, NoTriangleProof, AHCKGravity}.  In
ref.~\cite{NoTri} one-loop cancellations in generic theories of
gravity were linked to unexpectedly soft behavior of tree-level
gravity amplitudes under large complex shifts of their
momenta~\cite{GravityRecursion, AHK}.  This mechanism was also
proposed as a source of all-loop cancellations, perhaps sufficiently
strong, in conjunction with supersymmetric cancellations, to render
$\NeqEight$ supergravity ultraviolet finite.  This improved behavior
is unaccounted for in superspace power counting.

An important question is whether any of the known symmetries of
$\NeqEight$ supergravity can be used to constrain its multiloop
ultraviolet behavior.  For example, $\NeqEight$ supergravity contains
a non-compact $E_{7(7)}$ duality symmetry~\cite{CremmerJuliaScherk,
E7Original}.  Its explicit action on fields and amplitudes has been
presented recently~\cite{E7Recent,AHCKGravity}.  Improved ultraviolet
properties in $\NeqEight$ supergravity may also be linked to M-theory
dualities~\cite{DualityArguments} and to string theory
non-renormalization theorems~\cite{Berkovits,GreenII}, though issues
with docoupling of massive states~\cite{GOS} remain to be clarified.

A next step in unraveling the UV properties of $\NeqEight$
supergravity will be the calculation of the four-loop four-point
amplitude.  We have every reason to believe the four-loop amplitude
will be finite in four dimensions, but crucial guiding information
will come from identifying the smallest dimension in which the
divergence appears. This will allow us to test whether the observed
cancellations at three loops might have an explanation solely in
supersymmetry, along the lines of
refs.~\cite{HoweStelleNew,HoweStelleRecent}.  In particular, recent
supersymmetry arguments predict a four-loop divergence in $\NeqEight$
supergravity analytically continued to five
dimensions~\cite{HoweStelleRecent}.  On the other hand, generalized
unitarity arguments~\cite{Finite} point to $\NeqEight$ supergravity
being no more divergent than $\NeqFour$ super-Yang-Mills, which does
not have a four-loop five-dimensional divergence.  A four-loop
four-point calculation currently in progress should be able to resolve
this decisively~\cite{FourLoopYM}.

A key open problem is to develop a physical interpretation of the
observed ultraviolet cancellations in gravity theories and whether
this can be used to prove the ultraviolet finiteness of $\NeqEight$
supergravity.  More generally, it would be important to explore the
surprising and profound relations between gravity and gauge theories.
In particular, the KLT relations~\cite{KLT,Grant,FreitasGravity} and
their recent clarification~\cite{Twist} hint at a unification between
field theories of gravity and gauge theory of the sort implied by
string theory.


\section*{Acknowledgments}
\vskip -.3 cm 

We thank Lance Dixon, Darren Forde, Harald Ita, David Kosower and Radu
Roiban for many helpful discussions and collaborations on the results
summarized here.  This work was supported by the US Department of
Energy under contract DE--FG03--91ER40662.  J.~J.~M.~C. and
H.~J. gratefully acknowledge the financial support of Guy Weyl Physics
and Astronomy Alumni Fellowships.

\baselineskip 15pt

\parskip=.1cm


\end{document}